\documentclass[useAMS,usenatbib,usegraphicx]{mn2e}

\usepackage{times}
\usepackage{indentfirst}
\usepackage{amssymb}

\newcommand{\D}[2]{\frac{\partial #2}{\partial #1}}
\newcommand\bb[1] {\mbox{\boldmath{$#1$}}}
\newcommand\del{\bb{\nabla}} 
\newcommand\bcdot{\bb{\cdot}}
\newcommand\btimes{\bb{\times}} 

\title[Shear alpha-viscosity vs. turbulent magnetorotational stresses] {The
fundamental difference between shear alpha-viscosity and turbulent
magnetorotational stresses}

\author[M.E. Pessah, C.K. Chan, and D. Psaltis]
{Martin E. Pessah$^{1,3,4}$\thanks{E-mail:mpessah@ias.edu (MEP)},
Chi-kwan Chan$^{2,4}$, and Dimitrios Psaltis$^{3,4}$\vspace{6pt}\\
$^{1}$Institute for Advanced Study, School of Natural Sciences, Einstein Drive, Princeton, NJ, 08540, USA\\ 
$^{2}$ Institute for Theory and Computation, Harvard-Smithsonian Center for Astrophysics, 
60 Garden Street, Cambridge, MA 02138, USA\\ 
$^{3}$Astronomy Department, 933 N. Cherry Ave., Tucson, AZ,85721, USA\\ 
$^{4}$Physics Department, 1118 E. $4^{th}$ St., Tucson, AZ, 85721, USA}

\begin{document}

\date{Accepted --- . Received --- ; 
in original form --- }

\pagerange{\pageref{firstpage}--\pageref{lastpage}} \pubyear{2006}

\maketitle

\label{firstpage}

\begin{abstract}
  Numerical simulations of turbulent, magnetized, differentially
  rotating flows driven by the magnetorotational instability are often
  used to calculate the effective values of alpha viscosity that is
  invoked in analytical models of accretion discs. In this paper we
  use various dynamical models of turbulent magnetohydrodynamic
  stresses, as well as numerical simulations of shearing boxes, to show
  that angular momentum transport in MRI-driven accretion discs cannot be
  described by the standard model for shear viscosity. In
  particular, we demonstrate that turbulent magnetorotational stresses
  are not linearly proportional to the local shear and vanish
  identically for angular velocity profiles that increase outwards.
\end{abstract}



\section{Introduction}
\label{sec:intro}

It has long been recognized that molecular viscosity cannot be solely
responsible for angular momentum transport in accretion discs.
\citet{SS73} offered an appealing solution to this problem by
postulating a source of enhanced disc viscosity due to turbulence and
magnetic fields.  The standard accretion disc model rests on the idea
that the stresses between adjacent disc annuli are proportional to the
local shear, as in a Newtonian laminar shear flow, but that it is the
interaction of large turbulent eddies that results in efficient
transport.  The idea that turbulent angular momentum transfer in
accretion discs can be described in terms of an enhanced version of
the molecular transport operating in (laminar) differentially rotating
media has been at the core of the majority of studies in accretion
disc theory and phenomenology ever since \citep[see, e.g.,][]{FKR02}.

The origin of the turbulence that leads to enhanced angular momentum
transport in accretion discs has been a matter of debate since the
work of \citet{SS73}.  The issue of whether hydrodynamic turbulence
can be generated and sustained in astrophysical discs, due to the
large Reynolds numbers involved, is currently a matter of renewed
interest \citep{AMN05, MAN05}.  However, this idea has long been
challenged by analytical \citep{RG92, BH06}, numerical \citep{SB96,
BHS96, BH97, HBW99}, and, more recently, experimental work
\citep{JBSG2006}.

During the last decade, it has become evident that the interplay
between turbulence and magnetic fields is at a more fundamental level
than originally conceived. There is now strong theoretical and
numerical evidence suggesting that the process driving the turbulence
in accretion discs is related to a magnetic instability that operates
in the presence of a radially decreasing angular velocity profile.
Since the appreciation of the relevance of this magnetorotational
instability (MRI) to accretion physics (Balbus \& Hawley 1991; see
also Balbus \& Hawley 1998 and Balbus 2003 for a more recent review),
a variety of local \citep{HGB95, HGB96, SHGB96, Brandenburg95,
  Brandenburg01, Sanoetal04} and global
\citep{Armitage98,H00,H01,HK01, SP01} numerical simulations have
revealed that its long-term evolution gives rise to a turbulent state
and provides a natural avenue for vigorous angular momentum transport.

The fact that the overall energetic properties of turbulent
magnetohydrodynamic (MHD) accretion discs are similar to those of
viscous accretion discs \citep{BP99} has lead to the notion that
angular momentum transport due to MRI-driven turbulence in rotating
shearing flows can be described in terms of the alpha model
proposed by \citet{SS73}. This, in turn, has motivated many efforts
aimed to computing effective alpha values from numerical simulations
\citep[see, e.g.,][and references therein]{Gammie98, Brandenburg98} in
order to use them in large scale analytical models of accretion discs.

In this paper, we address in detail how the transport of angular
momentum mediated by MHD turbulence depends on the magnitude of the
local disc shear and contrast this result to the standard model for
shear viscosity.  We find that one of the fundamental assumptions in
which the standard viscous disc model is based, i.e., that angular
momentum transport is linearly proportional to the local shear, is not
appropriate for describing turbulent MRI-driven accretion discs.

\section{Alpha Viscosity vs. Turbulent MHD Stresses}
\label{sec:alpha_vs_mhd}

The equation describing the dynamical evolution of the mean angular
momentum density of a fluid element, $\bar{l}$, in an axisymmetric,
turbulent MHD accretion disc with tangled magnetic fields is
\begin{equation}
\label{eq:angular_momentum_mean}
\frac{\partial \bar l}{\partial t} + \frac{1}{r}
\frac{\partial}{\partial r} (r \bar l \bar{v}_r) =  
- \frac{1}{r} \frac{\partial}{\partial r} (r^2 \bar{T}_{r\phi}) \,.
\end{equation}
Here, the over-bars denote properly averaged values \citep[see,
e.g.,][]{BH98, BP99}, $\bar{v}_r$ is the radial mean flow velocity,
and the quantity $\bar{T}_{r\phi}$ represents the total stress
\begin{equation}
\label{eq:stresses_total}
\bar{T}_{r\phi} \equiv \bar{R}_{r\phi} - \bar{M}_{r\phi} \,
\end{equation}
acting on a fluid element as a result of the correlated fluctuations
in the velocity and magnetic fields in the turbulent flow, i.e.,
\begin{eqnarray}
\label{eq:stresses_reynolds}
\bar R_{r\phi} &\equiv& \langle \rho \delta\!v_r \delta\!v_\phi
\rangle \,, \\ 
\label{eq:stresses_maxwell}
\bar M_{r\phi} &\equiv& \frac{\langle \delta\!B_r \delta\!B_\phi\rangle}{4\pi} \,,
\end{eqnarray}
where $\bar R_{r\phi}$ and $\bar M_{r\phi}$ stand for the
$r\phi$-components of the Reynolds and Maxwell stress tensors,
respectively.

Equation (\ref{eq:angular_momentum_mean}) highlights the all-important
role played by correlated velocity and magnetic-field fluctuations in
turbulent accretion discs; if they vanish, the mean angular momentum
density of a fluid element is conserved. In order for matter in the
bulk of the disc to accrete, i.e., to loose angular momentum, the sign
of the mean total stress must be positive.  Note that the potential
for correlated kinetic fluctuations to transport angular momentum in
unmagnetized discs is still present in the hydrodynamic version of
equation (\ref{eq:angular_momentum_mean}).  However, the dynamical
role played by the correlated velocity fluctuations in hydrodynamic
flows is radically different from the corresponding role played by
correlated velocity and magnetic field fluctuations in MHD flows, even
when the magnetic fields involved are weak \citep{BH97,BH98}.

In order to calculate the structure of an accretion disc for which
angular momentum flows according to equation
(\ref{eq:angular_momentum_mean}), it is necessary to obtain a closed
system of equations for the second-order correlations defining the
total stress $\bar{T}_{r\phi}$.  In this context, the original
proposal by Shakura and Sunyaev \citep[see also,][]{LB74} can be seen
as a simple closure scheme for the correlations defining the total
turbulent stress in terms of mean flow variables (e.g., the pressure).

The model for angular momentum transport on which the standard
accretion disc is based rests on two distinct assumptions.  First, it
is postulated that the vertically averaged stress exerted on any given
disc annulus can be modeled as a shear viscous stress \citep{LB74},
i.e., in cylindrical coordinates,
\begin{equation}
\label{eq:shear_stress_turb}
\bar{T}^{\rm v}_{r\phi}  \equiv - r \Sigma \nu_{\rm turb} \frac{d\Omega}{dr} \,, 
\end{equation}
where $\Sigma$ and $\Omega$ stand for the vertically integrated disc
density and the angular velocity at the radius $r$, respectively.
This is a modification of the Newtonian model for the viscous stress
between adjacent layers in a differentially rotating laminar flow
\citep{LL59}; in this case, the coefficient $\nu_{\rm turb}$
parametrizes the turbulent kinematic viscosity.  In this model, the
direction of angular momentum transport is always opposite to the
angular velocity gradient.  This is the essence of a shear-driven
viscous disc.

Second, on dimensional grounds, it is assumed that the viscosity
coefficient can be parametrized as $\nu_{\rm turb} \equiv \alpha
c_{\rm s} H$.  This is because the physical mechanism that allows for
angular momentum transport is envisaged as the result of the
interaction of turbulent eddies of typical size equal to the disc
scale-height, $H$, on a turnover time of the order of $H/c_{\rm s}$,
where $c_{\rm s}$ is the isothermal sound speed. The parameter alpha
is often assumed to be constant and smaller than unity.  In the
presence of a shear background, the vertically integrated stress is
then given by
\begin{equation}
\label{eq:stress_standard}
\bar{T}^{\rm v}_{r\phi}  = - \alpha \bar{P} \frac{d\ln \Omega}{d\ln r} \,,
\end{equation}
where $\bar{P} = \Sigma c_{\rm s}^2/\gamma$ stands for the average pressure,
$\gamma$ is the ratio of specific heats, and we have used the fact
that the scale-height of a thin disc in vertical hydrostatic
equilibrium is roughly $H \sim c_{\rm s}/\Omega$.  This
parametrization of the coefficient of turbulent kinematic viscosity
implies that the efficiency with which angular momentum is transported
is proportional to the local average pressure.  This is the idea
behind the standard alpha-disc model.

It is worth mentioning that the expression usually employed to define
the stress in alpha-models, i.e., $\bar{T}^{\rm v}_{r\phi} = \alpha
\bar{P}$, only provides the correct order of magnitude for the stress
in terms of the pressure for a Keplerian discs. Indeed, in this case,
the shear parameter
\begin{equation}
\label{eq:shear_q}
q \equiv - \frac{d\ln \Omega}{d\ln r} \,,
\end{equation}
is equal to $3/2 \sim 1$.  However, the fact that the viscous stress
is proportional to the local shear cannot be overlooked in regions of
the disc where the local angular velocity can differ significantly
from its Keplerian value\footnote{Also note that it is the explicit
  linear dependence of the stress on the local shear what gives the
  equation for angular momentum transport its diffusive nature in
  standard accretion disc models.}.  This is expected to be the case
in the boundary layer around an accreting star or close to the
marginally stable orbit around a black hole.  These inner disc regions
are locally characterized by different, and possibly negative,
shearing parameters $q$.  More importantly, if the explicit dependence
of the stress on the local shear is not considered then equation
(\ref{eq:stress_standard}) predicts unphysical, non-vanishing stresses
for solid body rotation, $q\equiv 0$ \citep[c.f.,][]{Blaes04}.

More than a decade after the paper by \citet{BH91}, the MRI stands as
the most promising driver of the turbulence thought to enable the
accretion process. The stresses associated with this MRI-driven
turbulence have long been considered as the physical mechanism behind
the enhanced turbulent viscosity postulated in the standard model for
shear-driven angular momentum transport. It is important to note,
however, that there is no a priori reason to assume that the
correlated fluctuations defining the total turbulent stress in
MRI-driven turbulence are linearly proportional to the local shear. In
fact, this assumption can be challenged on both theoretical and
numerical grounds.


\section{Predictions from Stress Modelling}
\label{sec:stress_modelling}

There are currently few models that aim to describe the local dynamics
of turbulent stresses in differentially rotating magnetized media by
means of high-order closure schemes \citep{KY93, KY95, Ogilvie03,
PCP06b}\footnote{The studies described here concern models for the
evolution of the various correlated fluctuations relevant for
describing the dynamics of a turbulent magnetized flow.  We note that
\citet{VB97} proposed an incoherent dynamo model for the transport of
angular momentum driven by the generation of large-scale radial and
toroidal magnetic fields.}.  In these models, the total stress,
$\bar{T}_{r\phi}=\bar{R}_{r\phi}-\bar{M}_{r\phi}$, is not prescribed,
as in equation (\ref{eq:stress_standard}), but its value is calculated
considering the local energetics of the turbulent flow.  This is
achieved by deriving a set of non-linear coupled equations for the
various components of the Reynolds and Maxwell stress tensors.  These
equations involve unknown triple-order correlations among fluctuations
making necessary the addition of ad hoc closure relations.

Although the available models differ on the underlying physical
mechanisms that drive the turbulence and lead to saturation, some
important characteristics of the steady flows that they describe are
qualitatively similar. In particular, all of the models predict that
turbulent kinetic/magnetic cells in magnetized Keplerian discs are
elongated along the radial/azimuthal direction, i.e., $\bar R_{rr} >
\bar R_{\phi\phi}$ while $\bar M_{\phi\phi} > \bar M_{rr}$.
Furthermore, turbulent angular momentum transport is mainly carried by
correlated magnetic fluctuations, rather than by their kinetic
counterpart, i.e., $-\bar M_{r\phi} > \bar R_{r\phi}$.  All of these
properties are in general agreement with local numerical simulations.
However, a distinctive quantitative feature of these models that
concerns us here is that they make rather different predictions for
how the stresses depend on the magnitude and sign of the angular
velocity gradient.

In the remaining of this section, we highlight the most relevant
physical properties characterizing the various models and assess the
functional dependence of the total stress, $\bar T_{r\phi}$, on the
sign and steepness of the local angular velocity profile.  For
convenience, we summarize in Appendix A the various sets of equations
that define each of the models.

\subsection{Kato \& Yoshizawa 1995's Model}

In a series of papers, \citet{KY93, KY95} developed a model for
hydromagnetic turbulence in accretion discs with no large scale
magnetic fields \citep[see][for a review, and Nakao 1997 for the
inclusion of large scale radial and toroidal fields]{KFM98}. In their
closure scheme, triple-order correlations among fluctuations in the
velocity and magnetic fields are modeled in terms of second-order
correlations using the two-scale direct interaction approximation
\citep{Yoshizawa85, YII03}, as well as mixing length concepts.

In the model described in \citet{KY95}, the shear parameter $q$
appears explicitly only in the terms that drive the algebraic growth
of the turbulent stresses.  The physical mechanism that allows the
stresses to grow is the shearing of magnetic field lines.  This can be
seen by ignoring all the terms that connect the dynamical evolution of
the Reynolds and Maxwell stresses.  When this is the case, the
dynamics of these two quantities are decoupled. The Maxwell stress
exhibits algebraic growth while the Reynolds stress oscillates if the
flow is Rayleigh stable (i.e., $q<2$).  The growth of the magnetic
stresses is communicated to the different components of the Reynolds
stress via the tensor $\bar S_{ij} \equiv C_1^S \bar R_{ij} - C_2^S
\bar M_{ij}$, where $C_1^S$ and $C_2^S$ are model constants.

A characteristic feature of this model is that the physical mechanism
that leads to saturation is conceived as the escape of magnetic energy
in the vertical direction. This process is incorporated
phenomenologically by accounting for the leakage of magnetic energy
with terms of the form $-\beta \bar{M}_{ij}$, where $\beta$ stands for
the escape rate. Although turbulent kinetic and magnetic dissipation
act as sink terms in the equations for the various stress components
(proportional to the model parameters $\epsilon_{\rm G}$ and
$\epsilon_{\rm M}$, respectively), if it were not for the terms
accounting for magnetic energy escape, the system of equations would
be linear. This means that, in order for a stable steady-state
solution to exist, either the initial conditions or the constants
defining the model will have to be fine-tuned.

\begin{figure}
  \includegraphics[width=\columnwidth,trim=0 10 0 10]{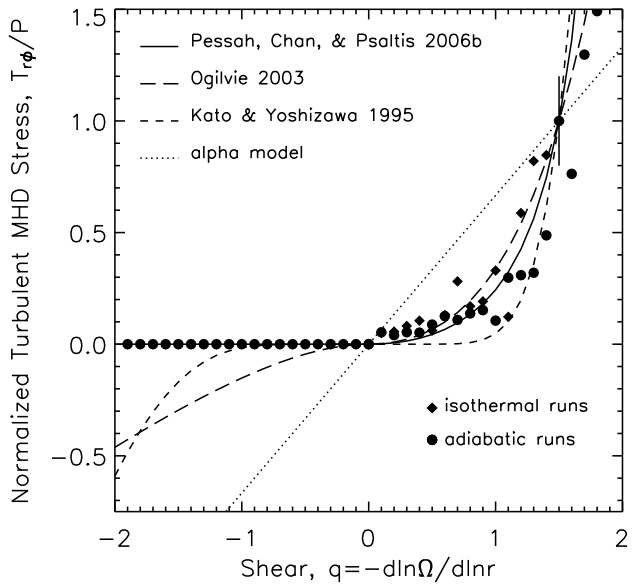}
  \caption{Total dimensionless turbulent stress at saturation as a
    function of the shear parameter $q\equiv-d\!\ln \Omega/d\!\ln r$.
    The various lines show the predictions corresponding to the three
    models for turbulent MHD angular momentum transport discussed in
    \S \ref{sec:stress_modelling}. The data points correspond to the
    volume and time averaged MHD stresses calculated from the series
    of shearing box simulations described in \S
    \ref{sec:stress_simulations}. The diamonds and circles represent
    the results obtained in the isothermal ($\gamma=1$) and adiabatic
    ($\gamma=5/3$) cases, respectively. The vertical bar in the
    simulations corresponding to the shear parameter $q=3/2$ shows the
    rms spread in the stresses as calculated from ten numerical
    simulations for a Keplerian disc. This spread, of roughly $20\%$,
    can be taken as a representative value for the typical rms spread
    in runs with different values of the parameter $q$ for both the
    isothermal and adiabatic cases.  The dotted line corresponds to a
    linear relationship between the stresses and the local shear, like
    the one assumed in the standard model for viscous angular momentum
    transport.  All the quantities in the figure are normalized to
    unity for a Keplerian profile, i.e., for $q=3/2$.}
  \label{fig:stress_shear}
\end{figure}

The functional dependence of the total stress on the shear parameter
$q$ for the model proposed in \citet{KY95} is shown in
Figure~\ref{fig:stress_shear}. The model predicts vanishingly small
stresses for small values of $|q|$; the total stress $\bar T_{r\phi}$
is a very strong function of the parameter $q$. Indeed, for $q>0$, the
model predicts $\bar T_{r\phi} \sim q^{8}$.


\subsection{Ogilvie 2003's Model}

Based on a set of fundamental principles constraining the non-linear
dynamics of turbulent flows, \citet{Ogilvie03} developed a model for
the dynamical evolution of the Maxwell and Reynolds stresses.  Five
non-linear terms accounting for key physical processes in turbulent
media are modeled by considering the form of the corresponding
triple-order correlations, energy conservation, and other relevant
symmetries.

The resulting model describes the development of turbulence in
hydrodynamic as well as in magnetized non-rotating flows.  An
interesting feature is that, depending on the values of some of the
model constants, it can develop steady hydrodynamic turbulence in
rotating shearing flows. For differentially rotating magnetized flows
with no mean magnetic fields, the physical mechanism that allows the
stresses to grow algebraically is the shearing of magnetic field
lines, as in the case of \citet{KY95}.  The transfer of energy between
turbulent kinetic and magnetic field fluctuations is mediated by the
tensor $c_3 \bar M^{1/2}\bar M_{ij} - c_4 \bar R^{-1/2} \bar M \bar
R_{ij}$, where $\bar R$ and $\bar M$ are the traces of the Reynolds
and Maxwell stresses, respectively, and $c_3$ and $c_4$ are model
constants. Note that, in this case, the terms that lead to
communication between the different Reynolds and Maxwell stress
components are intrinsically non-linear. The terms leading to
saturation are associated with the turbulent dissipation of kinetic
and magnetic energy and are given by $-c_1 \bar{R}^{1/2} \bar R_{ij}$
and $-c_5 \bar{M}^{1/2} \bar M_{ij}$, respectively.

The dependence of the total stress on the shear parameter for the
model described in \citet{Ogilvie03} is shown in
Figure~\ref{fig:stress_shear}.  For angular velocity profiles
satisfying $0<q<2$, the stress behaves as $\bar T_{r\phi} \sim q^{n}$
with $n$ between roughly 2 and 3.  Although the functional dependence
of the total stress on negative values of the shear parameter $q$ is
different from the one predicted by the model described in
\citet{KY95}, this model also generates MHD turbulence characterized
by negative stresses for angular velocity profiles increasing
outwards.

\subsection{Pessah, Chan, \& Psaltis 2006's Model}

Motivated by the similarities exhibited by the linear regime of the
MRI and the fully developed turbulent state \citep{PCP06a}, we have
recently developed a local model for the growth and saturation of the
Reynolds and Maxwell stresses in turbulent flows driven by the
magnetorotational instability \citep{PCP06b}.  Using the fact that the
modes with vertical wave-vectors dominate the fast growth driven by
the MRI, we obtained a set of equations to describe the exponential
growth of the different stress components.  By proposing a simple
phenomenological model for the triple-order correlations that lead to
the saturated turbulent state, we showed that the steady-state limit
of the model describes successfully the correlations among stresses
found in numerical simulations of shearing boxes \citep{HGB95}.

In the model described in \citet{PCP06b}, a new set of correlations
couple the dynamical evolution of the Reynolds and Maxwell stresses
and play a key role in developing and sustaining the magnetorotational
turbulence.  In contrast to the two previous cases, the tensor
connecting the dynamics of the Reynolds and Maxwell stresses cannot be
written in terms of $\bar R_{ij}$ or $\bar M_{ij}$. This makes it
necessary to incorporate additional dynamical equations for these new
correlations. In this model, all the second-order correlations exhibit
exponential (as opposed to algebraic) growth for shear parameters
$0<q<2$, in agreement with numerical simulations. Incidentally, this
is the only case in which the shear parameter, $q$, plays an explicit
role in connecting the dynamics of the Reynolds and Maxwell stresses.
The terms that lead to non-linear saturation are proportional to
$-(\bar M/\bar M_0)^{1/2}$, where $\bar M$ is the trace of the Maxwell
stress and $\bar M_0$ is a characteristic energy density set by the
local disc properties.

The functional dependence of the total turbulent stress on the local
shear for the model developed in \citet{PCP06b} is shown in
Figure~\ref{fig:stress_shear}.  For angular velocity profiles
decreasing outwards, i.e., for $q>0$, the stress behaves like $\bar
T_{r\phi} \sim q^{n}$ with $n$ between roughly 3 and 4. For angular
velocity profiles increasing outwards, i.e., for $q<0$, the stress
vanishes identically. Note that this is the only model that is
characterized by the absence of transport for all negative values of
the shear parameter $q$.

It is worth mentioning that in all three high-order closure schemes
described above, the pressure, $\bar{P}$, does not appear explicitly
in the equations defining the models. However, it does play a role in
setting the overall scale at which the stresses saturate.  This is
because the pressure provides a characteristic velocity (e.g., the
sound speed) or a characteristic length (e.g., the disc scale height)
which in turn determine the saturation level of the stresses
$\bar{T}_{r\phi}$.  In order to compare the predictions of how the
dimensionless stresses depend on the local shear independently of
other factors, we normalized the quantity $\bar{T}_{r\phi}/\bar{P}$
predicted by each model in Figure~\ref{fig:stress_shear} with the
values corresponding to the Keplerian cases with the same pressure
\footnote{Note that the model described in \citet{PCP06b} was
  developed to account for the correlations among stresses and
  magnetic energy density at saturation when there is a weak mean
  magnetic field perpendicular to the disc mid-plane.  It is known
  that, for a given initial magnetic energy density, numerical
  simulations of MHD turbulent Keplerian shearing flows tend to
  saturate at higher magnetic energy densities when there is a net
  vertical magnetic flux \citep{HGB96}.  The normalization chosen in
  Fig. \ref{fig:stress_shear} allows us to compare the
  shear-dependence of the various models regardless of their initial
  field configurations.}.

Figure~\ref{fig:stress_shear} illustrates the sharp contrast between
the functional dependence of the saturated stresses predicted by all
three MHD models with respect to the standard shear viscous stress
defined in equation (\ref{eq:shear_stress_turb}).  It is remarkable
that, despite the fact that the various models differ on their
detailed structure, all of them predict a much steeper functional
dependence of the stresses on the local shear. Indeed, for angular
velocity profiles decreasing outwards, they all imply $\bar{T}_{r\phi}
\sim q^n$ with $n\gtrsim 2$.  The predictions of the various models
differ more significantly for angular velocity profiles increasing
outwards. In this case, the models developed in \citet{KY95} and
\citet{Ogilvie03} lead to negative stresses, while the model developed
in \citet{PCP06b} predicts vanishing stresses.

\section{Results from Numerical Simulations}
\label{sec:stress_simulations}


There have been only few numerical studies to assess the properties of
magnetorotational turbulence for different values of the local shear.
\citet{ABL96} carried out a series of numerical simulations employing
the shearing box approximation to investigate the dependence of
turbulent magnetorotational stresses on the shear-to-vorticity ratio.
Although the number of angular velocity profiles that they considered
was limited, their results suggest that the relationship between the
turbulent MHD stresses and the shear is not linear. On the other hand,
\citet{HBW99} carried out a series of shearing box simulations varying
the shear parameter from $q=0.1$ up to $q=1.9$ in steps of $\Delta q
=0.1$ and reported on the dependence of the Reynolds and Maxwell
stresses on the magnitude of the local shear but not on the
corresponding dependence of the total stress.

In order to investigate the dependence of MRI-driven turbulent
stresses on the sign and magnitude of the local shear, we modified a
version of \texttt{ZEUS-3D} to allow for angular velocity profiles
increasing outwards (i.e., characterized by shear parameters $q<0$).
\texttt{ZEUS} is a publicly available code and is based on an explicit
finite difference algorithm on a staggered mesh.  A detailed
description of this code can be found in \citet{SN92a, SN92b} and
\citet{SMN92}.

\subsection{The Shearing Box Approximation}

The shearing box approximation has proven to be fruitful in studying
the local characteristics of magnetorotational turbulence from both
the theoretical and numerical points of view.  The local nature of the
MRI allows us to concentrate on scales much smaller than the scale
height of the accretion disc, $H$, and regard the background flow as
essentially homogeneous in the vertical direction.

In order to obtain the equations describing the dynamics of a
compressible MHD fluid in the shearing box limit, we consider a small
box centered at the radius $r_0$ and orbiting the central object in
corotation with the disc at the local speed $\bb{v}_0 = r_0\,\Omega_0
\check{\bb{\phi}}$.  The shearing box approximation consists of a
first order expansion in $r-r_0$ of all the quantities characterizing
the flow. The goal of this expansion is to retain the most important
terms governing the dynamics of the MHD fluid in a locally Cartesian
coordinate system \citep[see, e.g.,][]{GX94, HGB95}. This is a good
approximation as long as the magnetic fields involved are subthermal
\citep{PP05}.  The resulting set of equations is then given by
\begin{eqnarray}
 \D{t}{\rho} + \nabla\cdot(\rho\,\bb{v}) & = & 0
 \label{eq:cont} \\
 \D{t}{\bb{v}} + \left(\bb{v}\bcdot\del\right)\bb{v} & = &
 - \frac{1}{\rho}\del\left(P + \frac{\bb{B}^2}{8\pi}\right)
 + \frac{(\bb{B}\bcdot\del)\bb{B}}{4\pi\rho}
 \nonumber \\ & &
 - 2\bb{\Omega}_0 \btimes \bb{v} \, + \, q \Omega^2_0\del(r-r_0)^2
 \label{eq:momentum} \\
 \D{t}{\bb{B}} & = & \nabla\times(\bb{v}\times\bb{B})
 \label{eq:induction} \\
 \D{t}{E} + \nabla\cdot (E\,\bb{v}) & = & - P \nabla\cdot\bb{v}
 \label{eq:energy}
\end{eqnarray}
where $\rho$ is the density, $\bb{v}$ is the velocity, $\bb{B}$ is the
magnetic field, $P$ is the gas pressure, and $E$ is the internal
energy density. In writing this set of equations, we have neglected
the vertical component of gravity and defined the local Cartesian
differential operator,
\begin{equation}
 \del \equiv
 \check{\bb{r}} \, \frac{\partial}{\partial r}  +
 \frac{\check{\bb{\phi}}}{r_0}\,\frac{\partial}{\partial \phi} +
 \check{\bb{z}} \, \frac{\partial}{\partial z} \,,
\end{equation}
where $\check{\bb{r}}$, $\check{\bb{\phi}}$, and $\check{\bb{z}}$ are,
coordinate-independent, orthonormal basis vectors corotating with the
background flow at $r_0$.  Note that the third and fourth terms on the
right hand side of equation (\ref{eq:momentum}) account for the
Coriolis force, present in the rotating frame, and the radial
component of the tidal field, respectively. We close the set of
equations (\ref{eq:cont})--(\ref{eq:energy}) by assuming an ideal gas
law with $P=(\gamma-1)E$.

\subsection{Numerical Set Up}

We set the radial, azimuthal, and vertical dimensions of the
simulation domain to $L_r = 1$, $L_\phi = 6$, and $L_z = 1$ and
consider a grid of $32 \times 192 \times 32$ zones. This resolution
corresponds to the standard resolution used in most shearing box
simulations carried out up to date \citep[see, e.g.,][]{HGB95,Sanoetal04}.

The density scale in the shearing box is arbitrary and we choose
$\rho_0=1$ as in, e.g., \citet{HGB95} and \citet{Sanoetal04}.  We
consider the case of zero net magnetic flux through the vertical
boundaries\footnote{After submitting this paper we became aware of the
  impact of numerical resolution on the saturated stresses in
  Keplerian shearing boxes with zero-net flux.  \citet{PCP07}, see
  also \citet{FP07}, have shown that the saturation of the MRI in this
  type of simulations depends linearly on the numerical resolution.
  The numerical results in Figure~\ref{fig:stress_shear} are
  normalized with respect to the Keplerian case $(q=1.5)$. Note that
  the explicit functional dependence of the dimensionless stress on
  the shear, as shown in Figure~\ref{fig:stress_shear}, assumes that
  resolution affects all the simulations in the same way, regardless
  of the value of the shear $q$.} by defining the initial magnetic
field according to $\bb{B} = B_0 \sin[2\pi(r-r_0)/L_r]\check{\bb{z}}$.
The plasma $\beta$ in all the simulations that we perform is $\beta =
P/(B_0^2/8\pi) = 200$, so the magnetic field is highly subthermal in
the initial state.  The initial velocity field that corresponds to the
steady state solution is $\bb{v} = -q \Omega_0 (r-r_0)
\check{\bb{\phi}}$ and we choose the value $\Omega_0 = 10^{-3}$ in
order to set the time scale in the shearing box. Note that for $q =
3/2$, this velocity field is simply the first order expansion of a
steady Keplerian disc around $r_0$.  In order to excite the MRI, we
introduce random perturbations at the $0.1\%$ level in every grid
point over the background internal energy and velocity field in all of
the cases.

\subsection{Results}

Keeping all the numerical settings unchanged, we perform two suites of
numerical simulations with different values of the shear parameter
$q$, from $q=-1.9$ up to $q=1.9$ in steps of $\Delta q = 0.1$. The two
sets of runs differ only in the value of the adiabatic index $\gamma$;
we considered an isothermal case, with $\gamma = 1.001$, and an
adiabatic case, with $\gamma=5/3$.  For each value of the shear
parameter $q$, we run each simulation for 150 orbits.  We then compute
a statistically meaningful value for the saturated stress
$\bar{T}_{r\phi}$ and pressure $\bar{P}$ by averaging the last 100
orbits in each simulation \citep{WBH03, Sanoetal04}.

Figure~\ref{fig:stress_shear} shows the dimensionless stress
$\bar{T}_{r\phi}/\bar{P}$ obtained for both the isothermal and the
adiabatic cases (represented with diamonds and circles respectively)
as a function of the local shear.  It is evident from this figure that
the turbulent magnetorotational stresses are not proportional to the
local shear in either the isothermal or the adiabatic cases.  There is
indeed a strong contrast with respect to the standard assumption of a
linear relationship between the stresses and the local shear (dotted
line in the same figure) for both positive and negative shear
profiles.  For angular velocity profiles that decrease with increasing
radius, i.e., for $q>0$, all of the turbulent states are characterized
by positive mean stresses and thus by outward transport of angular
momentum.  In these cases, stronger shear results in larger saturated
stresses but the functional dependence of the total stress on the
local shear is not linear.  For angular velocity profiles that
increase with increasing radius, i.e., for $q<0$, all the numerical
simulations reach the same final state regardless of the magnitude of
the shear parameter $q$. The stresses resulting from the initial seed
perturbations (at the $0.1\%$ level) quickly decay to zero. This is in
sharp contrast with the large negative stresses that are implied by
the standard Newtonian model for the shear viscous stress in equations
(\ref{eq:shear_stress_turb}) and (\ref{eq:stress_standard}).

In order to explore further the decay of MHD turbulence found for
angular velocity profiles increasing outwards, we also performed the
following numerical experiment.  We seeded a run with a shear
parameter $q=-3/2$ with perturbations at the $100\%$ level of the
background internal energy and velocity field.  In this case, the
timescale for the decay of the initial turbulent state was longer than
the one observed in the corresponding run seeded with perturbations at
the $0.1\%$ level. The final outcome was nonetheless the same. After a
few orbits, the stresses decayed sharply and remained vanishingly
small until the end of the run at 150 orbits.  This result highlights
the strong stabilizing effects of a positive angular velocity gradient
on the dynamical evolution of the turbulent stresses due to tangled
magnetic fields.  This behavior can be understood in terms of the
joint restoring action due to magnetic tension and Coriolis forces
acting on fluid elements displaced form their initial orbits.

\section{Discussion and Implications}
\label{sec:discussion}

In this paper, we investigated the dependence of the turbulent
stresses responsible for angular momentum transport in differentially
rotating, magnetized media on the local shear as parametrized by
$q=-d\ln\Omega/d\ln r$. The motivation behind this effort lies in
understanding whether one of the fundamental assumptions on which much
of the standard accretion disc theory rests, i.e., the existence of a
linear relationship between these two quantities, holds when the MHD
turbulent state is driven by the MRI.  We addressed this problem both
in the context of current theoretical turbulent stress models as well
as using the publicly available three-dimensional numerical code
\texttt{ZEUS}.

From the theoretical point of view, we have seen that, despite their
different structures, all of the available high-order closure schemes
\citep{KY93, KY95, Ogilvie03, PCP06b} predict stresses whose
functional dependence on the local shear differ significantly from the
standard model for angular momentum transport.  In order to settle
this result on firmer grounds, we performed a series of numerical
simulations of MRI-driven turbulence in the shearing box approximation
for different values of the local shear characterizing the background
flow. The main conclusion to be drawn from our study is that turbulent
MHD stresses in differentially rotating flows are not linearly
proportional to the local background disc shear.  This finding
challenges one of the central assumptions in standard accretion disc
theory, i.e., that the total stress acting on a fluid element in a
turbulent magnetized disc can be modeled as a (Newtonian) viscous
shear stress.

We find that there is a strong contrast between the stresses produced
by MHD turbulence and the viscous shear stresses regardless of whether
the disc angular velocity decreases or increases outwards. In the
former case, i.e., for $q>0$ as in a Keplerian disc, the total
turbulent stress generated by tangled magnetic fields is not linearly
proportional to the local shear, $q$, as it is assumed in standard
accretion disc theory.  On the other hand, for angular velocity
profiles increasing outwards, i.e., for $q<0$ as in the boundary layer
close to a slowly rotating accreting stellar object, MHD turbulence
driven by the MRI fails to transport angular momentum, while viscous
shear stresses lead to enhanced negative stresses.

The functional dependence of the local stress on the shear profile
determines the topology of transonic accretion flows onto black holes
and the radial position of the corresponding critical points
\citep{AK89,KFM98,AP03}. It also plays a critical role in determining
the global structure of accretion flows onto stellar objects,
determining the exchange of angular momentum between the disc and the
gravitating body and even the angular velocity distribution itself
\citep{PN91}. If magnetorotational turbulence is the main mechanism
enabling angular momentum transport in accretion discs then the
dependence of the stress on the local shear discussed in this paper
can have important implications for the global structure and long term
evolution of accretion disc around proto-stars, proto-neutron stars,
accreting binaries, and active galactic nuclei.

\section*{Acknowledgments}
We are grateful to Jim Stone for useful discussions and for helping us
with the necessary modifications to the \texttt{ZEUS} code.  We thank
Gordon Ogilvie for his detailed comments on an earlier version of this
manuscript. We have also benefit with fruitful discussions with Eric
Blackman, Omer Blaes, Phil Armitage, and Andrew Cumming on different
aspects of stress modelling and accretion theory. We also thank an
anonymous referee for useful comments and constructive criticisms. MEP
was supported through a Jamieson Fellowship at the Astronomy
Department at the UA during part of this study.  This work was
partially supported by NASA grant NAG-513374.

\appendix

\section{High-order closure models}
\label{sec:appendix_a}

We summarize here the various sets of equations defining the models
described in \S\ref{sec:stress_modelling}. In order to simplify the
comparison between the different models, we adopt the notation
introduced in \S\ref{sec:alpha_vs_mhd}, even if this was not the
original notation used by the corresponding authors.  With the same
motivation, we work with dimensionless sets of equations obtained by
using the inverse of the local angular frequency ($\Omega_0^{-1}$) as
the unit of time and the relevant characteristics speeds or lengths
involved in each case. We also provide here the values of the various
model constants that were adopted in order to obtain the total
turbulent stresses as a function of the local shear shown in
Figure~\ref{fig:stress_shear}.

\subsection{Kato \& Yoshizawa 1995' s Model} 

The model defined by Kato \& Yoshizawa is based on the two-scale
direct interaction approximation \citep[][]{Yoshizawa85, YII03}. The
temporal evolution of the Reynolds and Maxwell stresses is described
by
\begin{eqnarray}
\label{eq:mean_Rrr_ky}
\partial_t \bar{R}_{rr} &=& 4 \bar{R}_{r\phi} + \bar{\Pi}_{rr} - \bar{S}_{rr} - \frac{2}{3} \epsilon_{\rm G}
\,,  \nonumber \\
\label{eq:mean_Rrphi_ky}
\partial_t \bar{R}_{r\phi} &=& (q-2) \bar{R}_{rr} + 2 \bar{R}_{\phi\phi} + \bar{\Pi}_{r\phi} - \bar{S}_{r\phi}
\,, \nonumber \\
\label{eq:mean_Rphiphi_ky}
\partial_t \bar{R}_{\phi\phi} &=& 2(q-2) \bar{R}_{r\phi}
+ \bar{\Pi}_{\phi\phi} - \bar{S}_{\phi\phi} - \frac{2}{3} \epsilon_{\rm G}
\,,  \nonumber \\ 
\label{eq:mean_Rzz_ky}
\partial_t \bar{R}_{zz} &=& 
\bar{\Pi}_{zz} - \bar{S}_{zz} - \frac{2}{3} \epsilon_{\rm G}
\,,  \nonumber \\ 
\label{eq:mean_Mrr_ky}
\partial_t \bar{M}_{rr} &=& \bar{S}_{rr} - 2 \beta\bar{M}_{rr} - \frac{2}{3} \epsilon_{\rm M}
\,, \nonumber \\
\label{eq:mean_Mrphi_ky}
\partial_t \bar{M}_{r\phi} &=& -q \bar{M}_{rr} + \bar{S}_{r\phi} - 2 \beta\bar{M}_{r\phi}
\,,  \nonumber \\
\label{eq:mean_Mphiphi_ky}
\partial_t \bar{M}_{\phi\phi} &=& -2q \bar{M}_{r\phi}
+ \bar{S}_{\phi\phi} -2 \beta\bar{M}_{\phi\phi} - \frac{2}{3} \epsilon_{\rm M}
\,, \nonumber \\
\label{eq:mean_Mzz_ky}
\partial_t \bar{M}_{zz} &=&  \bar{S}_{zz} - 2 \beta\bar{M}_{zz} - \frac{2}{3} \epsilon_{\rm M}
\,,  \nonumber
\end{eqnarray}
where the relevant characteristic speed used to define dimensionless
variables is the local sound speed.

In this model, the flow of turbulent energy from kinetic to magnetic
field fluctuations is determined by the tensor
\begin{eqnarray}
  \label{eq:S_ij_ky}
\bar{S}_{ij} = C_1^{\rm S} \bar{R}_{ij} - C_2^{\rm S}\bar{M}_{ij} \,,
\end{eqnarray}
where $C_1^S$ and $C_2^S$ are model constants. This quantity plays the
most relevant role in connecting the dynamics of the different
components of the Reynolds and Maxwell tensors.  If $\bar{S}_{ij}$ is
positive, the interactions between the turbulent fluid motions and
tangled magnetic fields enhances the latter.  The pressure-strain
tensor is modeled in the framework of the two scale direct interaction
approximation according to
\begin{eqnarray}
\label{eq:Piij_ky}
\bar{\Pi}_{ij} &=& 
  -   C_1^{\rm \Pi} (\bar{R}_{ij} - \delta_{ij}\bar R/3)
  -   C_2^{\rm \Pi} (\bar{M}_{ij} - \delta_{ij}\bar M/3) \nonumber \\
&&- q \, C_0^{\rm \Pi} \bar{R}\,(\delta_{ir}\delta_{j\phi}+\delta_{i\phi}\delta_{jr}) \,,
\end{eqnarray}
where $\bar{R}$ and $\bar{M}$ stand for the traces of the Reynolds and
Maxwell stresses, respectively.  This tensor accounts for the
redistribution of turbulent kinetic energy along the different
directions and tends to make the turbulence isotropic.  The
dissipation rates are estimated using mixing length arguments and are
modeled according to
\begin{eqnarray}
\label{eq:E_ky}
\epsilon_{\rm G} &=& \frac{3}{2}\nu_{\rm G}(\bar{R} + \bar{M}) \,, \\
\epsilon_{\rm M} &=& \frac{3}{2}\nu_{\rm M}(\bar{R} + \bar{M}) \,,
\end{eqnarray}
where $\nu_{\rm G}$ and $\nu_{\rm M}$ are dimensionless constants.
The escape of magnetic energy in the vertical direction is taken into
account phenomenologically via the terms proportional to the
(dimensionless) rate
$\beta = X \bar{M}^{1/2}$,
with $0<X<1$.

The values of the constants that we considered in order to obtain the
curve shown in Figure~\ref{fig:stress_shear} are the same as the ones
considered in case 2 in \citet{KY95}, i.e., $C_0^{\rm \Pi} = C_1^{\rm
\Pi} = C_2^{\rm \Pi} = C_1^{\rm S}= C_2^{\rm S}=0.3$, $\nu_{\rm
G}=\nu_{\rm M}=0.03$. We further consider $X=0.5$ as a representative
case.

\subsection{Ogilvie 2003's Model} 

Ogilvie proposed that the functional form of the equations governing
the local, non-linear dynamics of a turbulent MHD flow are strongly
constrained by a set of fundamental principles. The model that he
developed is given by the following set of equations
\begin{eqnarray}
\label{eq:mean_Rrr_nl}
\partial_t \bar{R}_{rr} &=& 4 \bar{R}_{r\phi} 
 -  c_1\bar{R}^{ 1/2} \bar{R}_{rr} - c_2\bar{R}^{ 1/2}(\bar{R}_{rr}-\bar R/3) \nonumber \\
&&+ c_3\bar{M}^{ 1/2} \bar{M}_{rr}- c_4\bar{R}^{-1/2} \bar{M}\bar{R}_{rr}
\,,  \nonumber\\
\label{eq:mean_Rrphi_nl}
\partial_t \bar{R}_{r\phi} &=& (q-2) \bar{R}_{rr} + 2 \bar{R}_{\phi\phi} 
 - (c_1 + c_2)\bar{R}^{ 1/2} \bar{R}_{r\phi}\nonumber \\
&&+ c_3\bar{M}^{ 1/2} \bar{M}_{r\phi}- c_4\bar{R}^{-1/2} \bar{M}\bar{R}_{r\phi}
\,,  \nonumber \\
\label{eq:mean_Rphiphi_nl}
\partial_t \bar{R}_{\phi\phi} &=& 2(q-2) \bar{R}_{r\phi}
 -  c_1\bar{R}^{ 1/2} \bar{R}_{\phi\phi} - c_2\bar{R}^{ 1/2}(\bar{R}_{\phi\phi}-\bar R/3) \nonumber \\
&&+ c_3\bar{M}^{ 1/2} \bar{M}_{\phi\phi}- c_4\bar{R}^{-1/2} \bar{M}\bar{R}_{\phi\phi}
\,,   \nonumber \\ 
\label{eq:mean_Rzz_nl}
\partial_t \bar{R}_{zz} &=& 
 -  c_1\bar{R}^{ 1/2} \bar{R}_{zz} - c_2\bar{R}^{ 1/2}(\bar{R}_{zz}-\bar R/3) \nonumber \\
&&+ c_3\bar{M}^{ 1/2} \bar{M}_{zz}- c_4\bar{R}^{-1/2} \bar{M}\bar{R}_{zz}
\,,  \nonumber  \\ 
\label{eq:mean_Mrr_nl}
\partial_t \bar{M}_{rr} &=& 
c_4\bar{R}^{-1/2}\bar{M}\bar{R}_{rr}-(c_3+c_5) \bar{M}^{1/2}\bar{M}_{rr}
\,,  \nonumber\\
\label{eq:mean_Mrphi_nl}
\partial_t \bar{M}_{r\phi} &=& - q \bar{M}_{rr}
 +  c_4\bar{R}^{-1/2}\bar{M}\bar{R}_{r\phi} 
-(c_3+c_5) \bar{M}^{1/2}\bar{M}_{r\phi}
\,, \nonumber \\
\label{eq:mean_Mphiphi_nl}
\partial_t \bar{M}_{\phi\phi} &=& -2q\bar{M}_{r\phi} 
+ c_4\bar{R}^{-1/2}\bar{M}\bar{R}_{\phi\phi} 
-(c_3+c_5) \bar{M}^{1/2}\bar{M}_{\phi\phi} \,,\nonumber \\
\label{eq:mean_Mzz_nl}
\partial_t \bar{M}_{zz} &=&
c_4\bar{R}^{-1/2}\bar{M}\bar{R}_{zz}-(c_3+c_5)\bar{M}^{1/2}\bar{M}_{zz}
\,.  \nonumber
\end{eqnarray}
Here $\bar{R}$ and $\bar{M}$ denote the traces of the Reynolds and
Maxwell tensors and we have defined the quantities $c_1, \dots, c_5$
which are related to the positive dimensionless constants defined by
Ogilvie $C_1, \dots, C_5$ via $C_i=L c_i$, where $L$ is a vertical
characteristic length (e.g., the thickness of the disc). Note that
Ogilvie's original equations are written in terms of Oort's first
constant $A =q/2$ (in dimensionless units).

In this model, the constant $c_2$ dictates the return to isotropy
expected to be exhibited by freely decaying hydrodynamic turbulence.
The terms proportional to $c_3$ and $c_4$ transfer energy between
kinetic and magnetic turbulent fields. The constants $c_1$ and $c_5$
are related to the dissipation of turbulent kinetic and magnetic
energy, respectively. Note that, in order to obtain the representative
behavior of the total turbulent stress as a function of the local
shear that is shown Figure~\ref{fig:stress_shear}, we set $c_1, \dots,
c_5 =1$.

\subsection{Pessah, Chan, \& Psaltis 2006's Model}

We have recently developed a local model for the growth and saturation
of the Reynolds and Maxwell stresses in turbulent flows driven by the
magnetorotational instability that leads to exponential growth for the
stresses and can account for a number of correlations observed in
numerical simulations \citep{PCP06b}.  In this model, the Reynolds and
Maxwell stresses are not only coupled by the same linear terms that
drive the turbulent state in the previous two models but there is also
a new tensorial quantity that couples their dynamics further. This new
tensor cannot be written in terms of $\bar R_{ij}$ or $\bar M_{ij}$,
making it necessary to incorporate additional dynamical equations.

The set of equations defining this model is
\begin{eqnarray}
\label{eq:mean_Rrr}
\partial_t \bar{R}_{rr} &=& 4 \bar{R}_{r\phi}+2
\bar{W}_{r\phi} -\sqrt{\frac{\bar{M}}{\bar{M}_0}} \bar{R}_{rr}
\,, \nonumber  \\
\label{eq:mean_Rrphi}
\partial_t \bar{R}_{r\phi} &=& 
(q-2) \bar{R}_{rr} + 2 \bar{R}_{\phi\phi} - \bar{W}_{rr} +
\bar{W}_{\phi\phi} -\sqrt{\frac{\bar{M}}{\bar{M}_0}} \bar{R}_{r\phi}
\,, \nonumber \\
\label{eq:mean_Rphiphi}
\partial_t \bar{R}_{\phi\phi} &=& 2(q-2) \bar{R}_{r\phi}
- 2 \bar{W}_{\phi r} -\sqrt{\frac{\bar{M}}{\bar{M}_0}}
\bar{R}_{\phi\phi}\,,  \nonumber \\ 
\label{eq:mean_Wrr}
\partial_t \bar{W}_{rr} \!\!&=&\!\! q \bar{W}_{r \phi} + 2 \bar{W}_{\phi r} + 
 \zeta^2 k^2_{\rm max} (\bar{R}_{r\phi} - \bar{M}_{r\phi}) 
- \sqrt{\frac{\bar{M}}{\bar{M}_0}} \bar{W}_{rr}\,,  \nonumber \\
\label{eq:mean_Wrphi}
\partial_t \bar{W}_{r\phi} \!\!&=&\!\! 2 \bar{W}_{\phi \phi} -  
 \zeta^2 k^2_{\rm max} (\bar{R}_{rr} -\bar{M}_{rr})
- \sqrt{\frac{\bar{M}}{\bar{M}_0}} \bar{W}_{r\phi}
\,,  \nonumber \\
\label{eq:mean_Wphir}
\partial_t \bar{W}_{\phi r} \!\!&=&\!\! (q-2) \bar{W}_{rr} 
+ q \bar{W}_{\phi\phi} \nonumber \\ 
&+&  \zeta^2 k^2_{\rm max}(\bar{R}_{\phi\phi} - \bar{M}_{\phi\phi})
- \sqrt{\frac{\bar{M}}{\bar{M}_0}} \bar{W}_{\phi r}
\,,  \nonumber \\
\label{eq:mean_Wphiphi}
\partial_t \bar{W}_{\phi\phi} \!\!&=&\!\! (q-2) \bar{W}_{r\phi} - 
 \zeta^2 k^2_{\rm max} (\bar{R}_{r\phi} - \bar{M}_{r\phi})
- \sqrt{\frac{\bar{M}}{\bar{M}_0}} \bar{W}_{\phi\phi}
\,, \nonumber \\
\label{eq:mean_Mrr}
\partial_t \bar{M}_{rr} &=& -2 \bar{W}_{r\phi} -\sqrt{\frac{\bar{M}}{\bar{M}_0}} \bar{M}_{rr}
\,, \nonumber  \\
\label{eq:mean_Mrphi}
\partial_t \bar{M}_{r\phi} &=& -q \bar{M}_{rr}
+ \bar{W}_{rr} - \bar{W}_{\phi\phi}-\sqrt{\frac{\bar{M}}{\bar{M}_0}} \bar{M}_{r\phi}
\,,  \nonumber \\
\label{eq:mean_Mphiphi}
\partial_t \bar{M}_{\phi\phi} &=& -2q \bar{M}_{r\phi}
+2 \bar{W}_{\phi r} -\sqrt{\frac{\bar{M}}{\bar{M}_0}} \bar{M}_{\phi\phi} \nonumber 
\,,
\end{eqnarray}
where we have defined dimensionless variables considering the mean
Alfv\'en speed $\bar{v}_{{\rm A} z}=\bar{B}_z/\sqrt{4\pi\rho_0}$, with
$\bar{B}_z$ the local mean magnetic field in the vertical direction
and $\rho_0$ the local disc density. The tensor
$\bar{W}_{ik}=\langle\delta v_i \delta j_k\rangle$ is defined in terms
of correlated fluctuations in the velocity and current ($ \delta
\bb{j} = \del \btimes \delta \bb{B}$) fields. The (dimensionless)
wavenumber defined as
\begin{eqnarray}
\label{eq:kbar}
k^2_{\rm max} = q - \frac{q^2}{4} 
\end{eqnarray}
corresponds to the scale at which the MRI-driven fluctuations
exhibit their maximum growth and 
\begin{equation}
\bar{M}_0\equiv \xi \rho_0 H\Omega_0  \bar{v}_{{\rm A} z} \,,
\label{eq:m0}
\end{equation}
is a characteristic energy density set by the local disc properties,
with $H$ the disc thickness.  The parameters $\zeta \simeq 0.3$ and
$\xi \simeq 11$ are model constants which are determined by requiring
that the Reynolds and Maxwell stresses satisfy the correlations
observed in numerical simulations of Keplerian shearing boxes with
$q=3/2$ \citep{HGB95}.


\label{lastpage}

\end{document}